\documentclass{caosp302}
\usepackage{graphicx,amsmath,amssymb}

\articleNo{123} \pubyear{2007} \volume{3} \volnumber{3}
\firstpage{1} \received{October 1, 2007} \accepted{October 1, 2005}

\begin{document}

\hauthor{Z.\,Mikul\'a\v sek, T. Gr\'af, J.\,Krti\v cka, J.\,Zverko,
and J.\,\v Zi\v z\v novsk\'y}

\htitle{Photometrically simply behaving mCP stars}

\title{Photometrically simply behaving mCP stars}

\author{Z. Mikul\'a\v sek \inst{1,2} \and T. Gr\'af \inst{2}\and
J. Krti\v cka \inst{1} \and J. Zverko \inst{3} \and J. \v Zi\v z\v
novsk\'y \inst{3}}

\institute{ Institute of Theoretical Physics and Astrophysics,
            Masaryk University, Brno, Czech Republic,
        \and
            J.\,Palisa Observatory and Planetarium, V\v SB TU,
            Ostrava, Czech Republic
        \and
            Astronomical Institute, Tatransk\'a Lomnica,
            Slovak Republic}

\date{November 28, 2002}

\maketitle

\begin{abstract}
We analyzed $ubvy$ and $H_{\rm p}$ light curves of 19 well observed
magnetic CP stars selected from the {\it On-line database of
photometric observations of mCP stars} of which light curves in all
the five colours were similar. We assumed that among these
photometrically simply behaving (PSB) stars could be found such ones
which have a single photometric spot. The insight into such simple
situations would help us to comprehend more complicated cases.

Light curves of the 19 PSB mCP stars proved to be generally nearly
symmetric but surprisingly diverse. The analysis shows that only in
the case of HD\,110956B, HD\,188041, and perhaps HD\,193722 we are
able to explain their photometric behaviour by a simple one-spot
model. Consequently, occurrence of more than one photometric spot on
an mCP star is typical.

\end{abstract}

\section{Introduction}

It is generally accepted that the rotationally modulated photometric
variability of magnetic CP stars originates in consequence of
non-uniform horizontal structure of their atmospheres induced by
strong global magnetic field and uneven surface distribution of
chemical elements.

Recently we successfully simulated the observed light variability of
the He-strong CP star HD\,37776 in $uvby$ colours providing an
inhomogeneous elemental horizontal distribution. Despite this star
displays a relatively complex surface pattern with several
relatively bright and/or dark regions (see Krti\v{c}ka {\it et al.}
2007) and a quadrupole dominated magnetic field (Thompson,
Landstreet 1985), its light curve shows a not very complicated
single wave form. However, HD\,37776 is not a typical representative
for its considerably complex magnetic field geometry which can make
the stellar surface a bit disordered. A still better subject could
be another Si star, HD\,177410, (Krti\v{c}ka {\it et al.} 2008,
these proceedings) with only two bright spots on the surface, and we
do not exclude the existence of mCP stars with even simpler
photometric structure having only one dominant spot tied with
spectroscopic or magnetic structures on their surfaces.

In this brief recognition study we attempt to look up and study mCP
stars which could carry on such easily readable surface structures.
The comprehension of the simplest case can be a way to common, more
complicated ones.

\section{Photometrically simply behaving mCP stars}

We showed (Mikul\'a\v{s}ek {\it et al.} 2007a) that the variable
part of a light curve (LCs), $\Delta m_c$~in the colour $c$ of an
mCP star, can be expressed as a linear combination of two normalized
orthogonal phase functions $f_1(\varphi),f_2(\varphi)$, which can be
found easily by means of the advanced principal component analysis
(Mikul\'a\v{s}ek, 2007):
\begin{eqnarray}
\Delta m_c(\varphi)\cong A_{1c}\,f_1(\varphi)+A_{2c}
f_2(\varphi);\label{pca}
\end{eqnarray}
\begin{eqnarray}
\sum_{j=1}^N f_1^2(\varphi_j)=\sum_{j=1}^N
f_2^2(\varphi_j)\cong\frac{N}{2};\quad \sum_{j=1}^N
f_1(\varphi_j)f_2(\varphi_j)=0,
\end{eqnarray}
where $N$ is the number of measurements, $\varphi_j$ is phase of
the $j$-th measurement.

We quantify the similarity of two colour light curves by a ratio
\begin{eqnarray}
 r^2=\sum_{c=1}^5 A_{2c}^2/\sum_{c=1}^5 A_{1c}^2,
\end{eqnarray}
and we adopt that light curves of an mCP star are mutually similar
if $r<0.15$. In such a degree of similarity the light curves
can be described sufficiently by a unique phase function, which
mathematically means the second term in Eq.\,\ref{pca} can be neglected.

\begin{figure}[t]
\centerline{\includegraphics[width=0.492\textwidth]{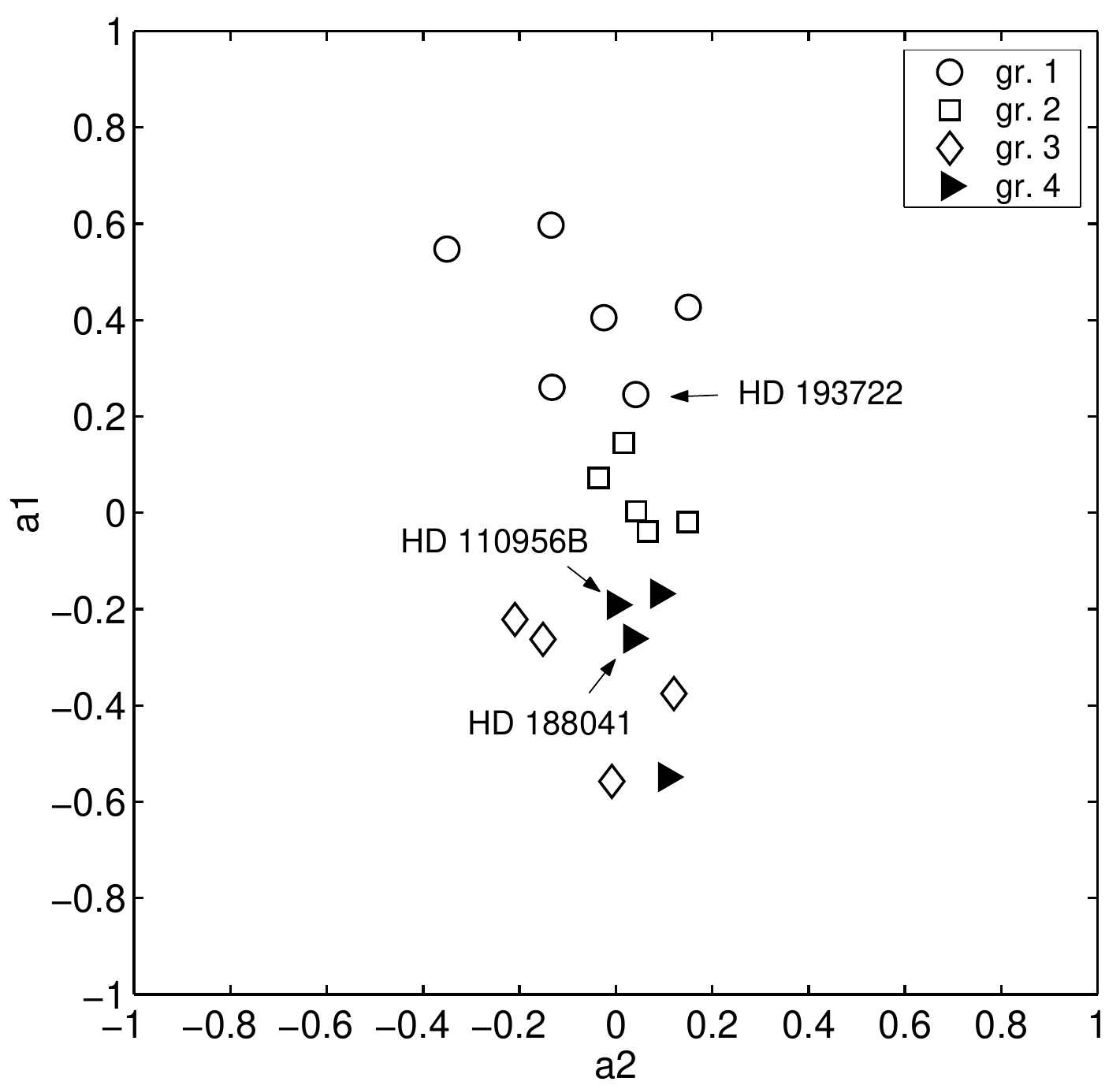}
\includegraphics[width=.500\textwidth,height=.47\textwidth]{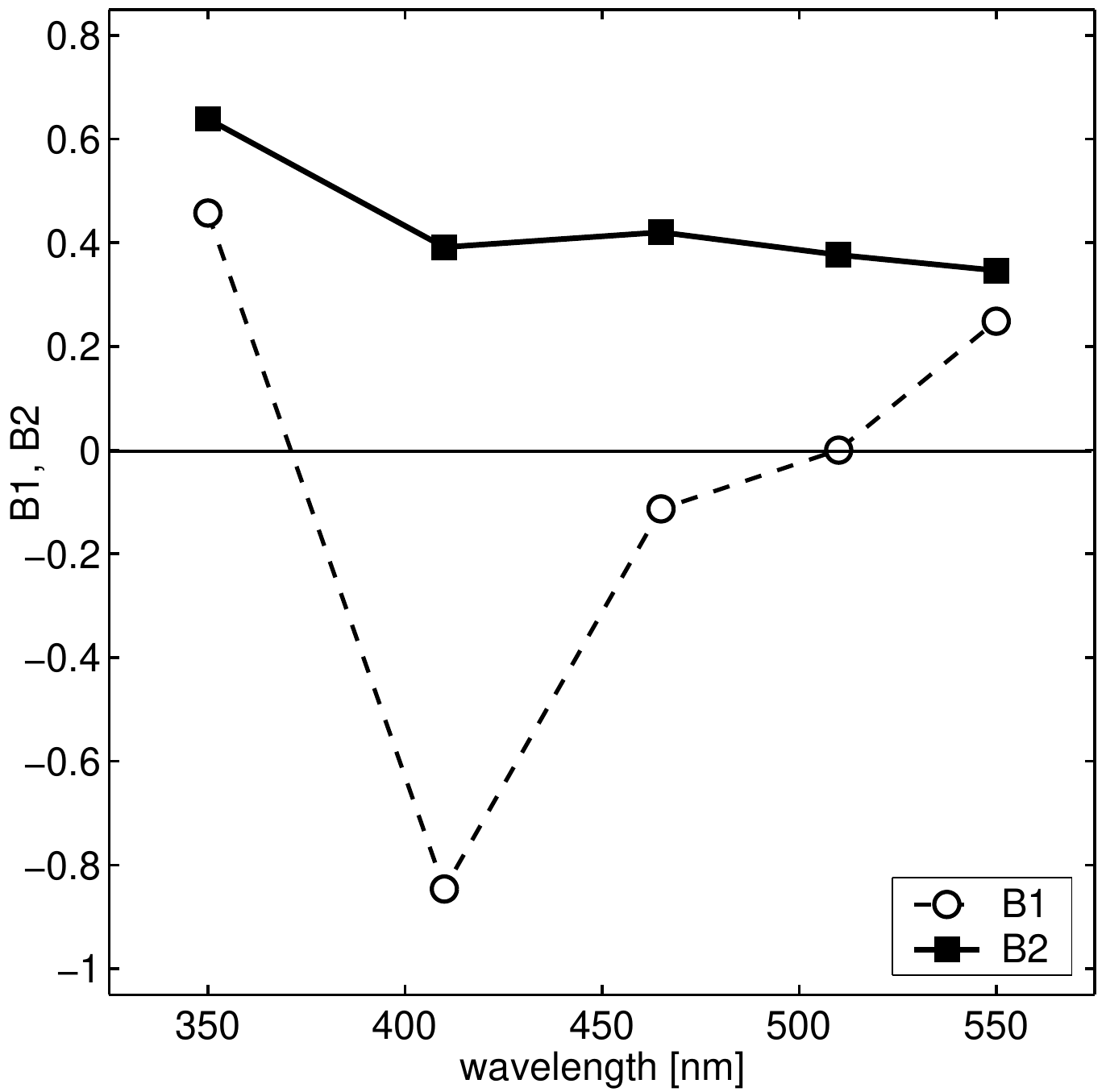}}
\caption{{\bf Left:} The relation of LCs' parameters $a_1$ and
$a_2$, shows that LCs of our PSB stars are only slightly asymmetric.
LCs can be divided into four groups. Three stars whose LC's can be
interpreted by the one-spot model are denoted with their HD names.
{\bf Right:} Two basic dependencies of LCs' amplitudes on the
effective wavelength.} \label{a1a2}
\end{figure}

\begin{figure}[tb]
\centering
\resizebox{0.3\hsize}{0.2\vsize}{\includegraphics{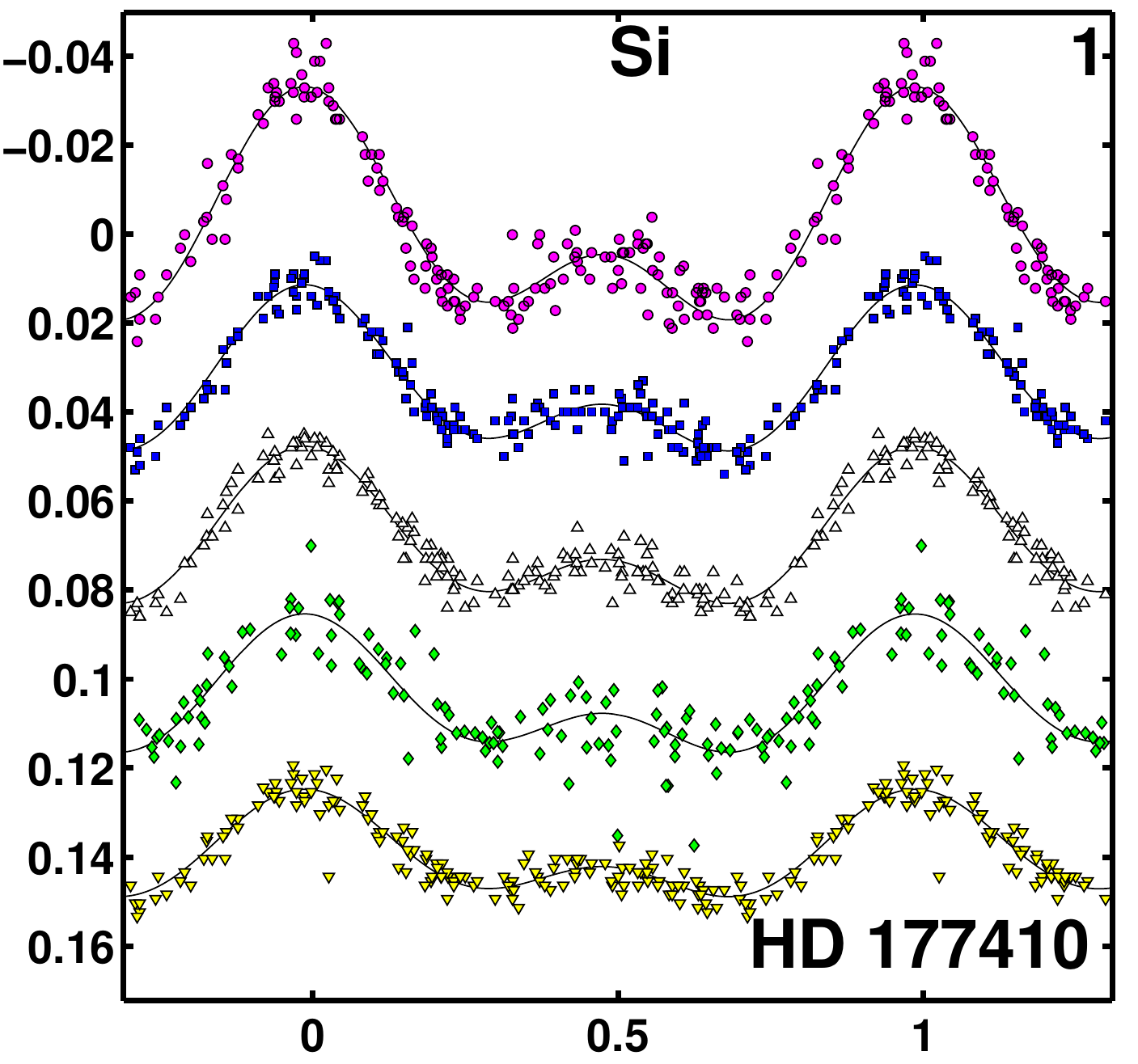}}
\resizebox{0.3\hsize}{0.2\vsize}{\includegraphics{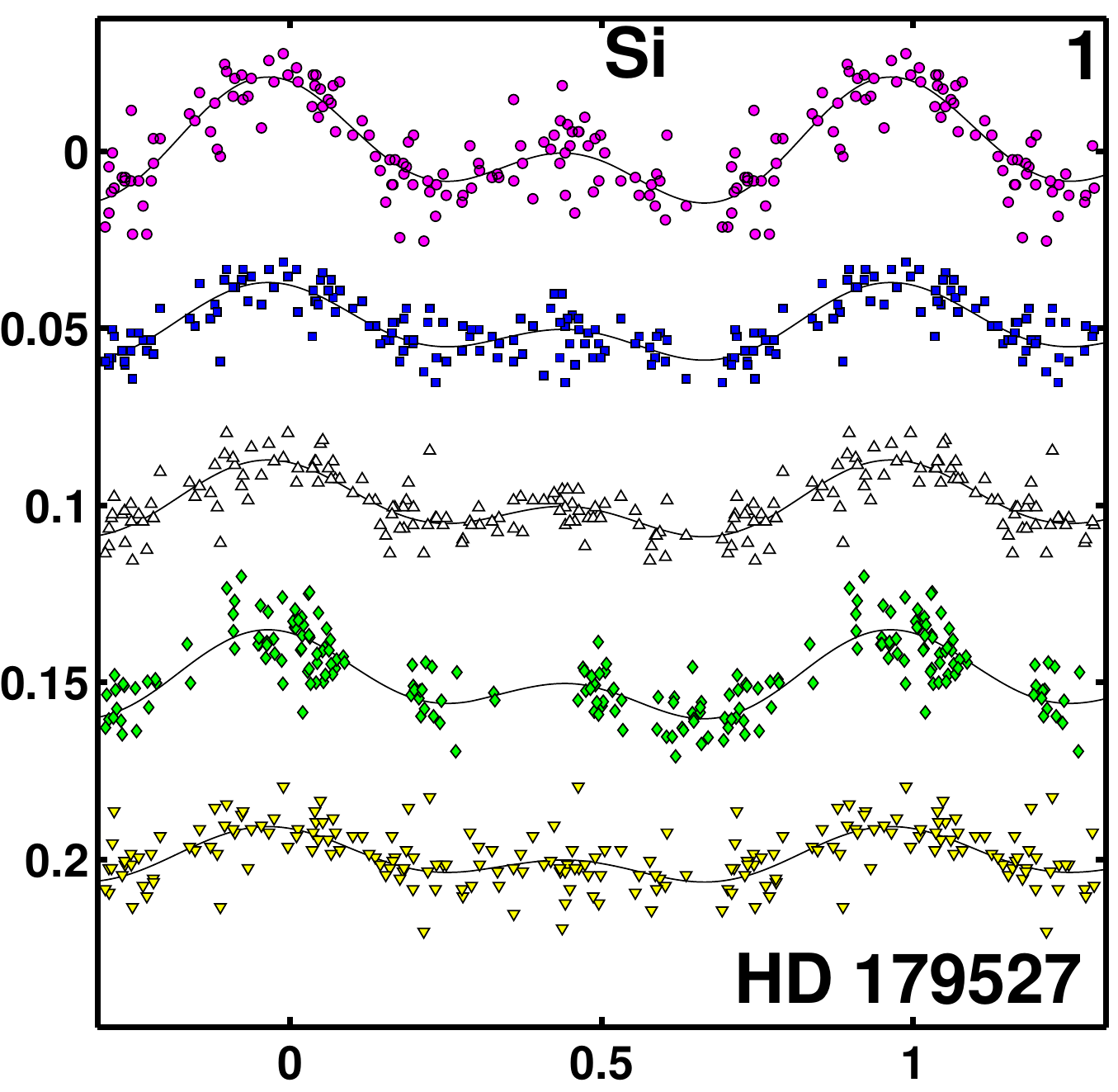}}
\resizebox{0.3\hsize}{0.2\vsize}{\includegraphics{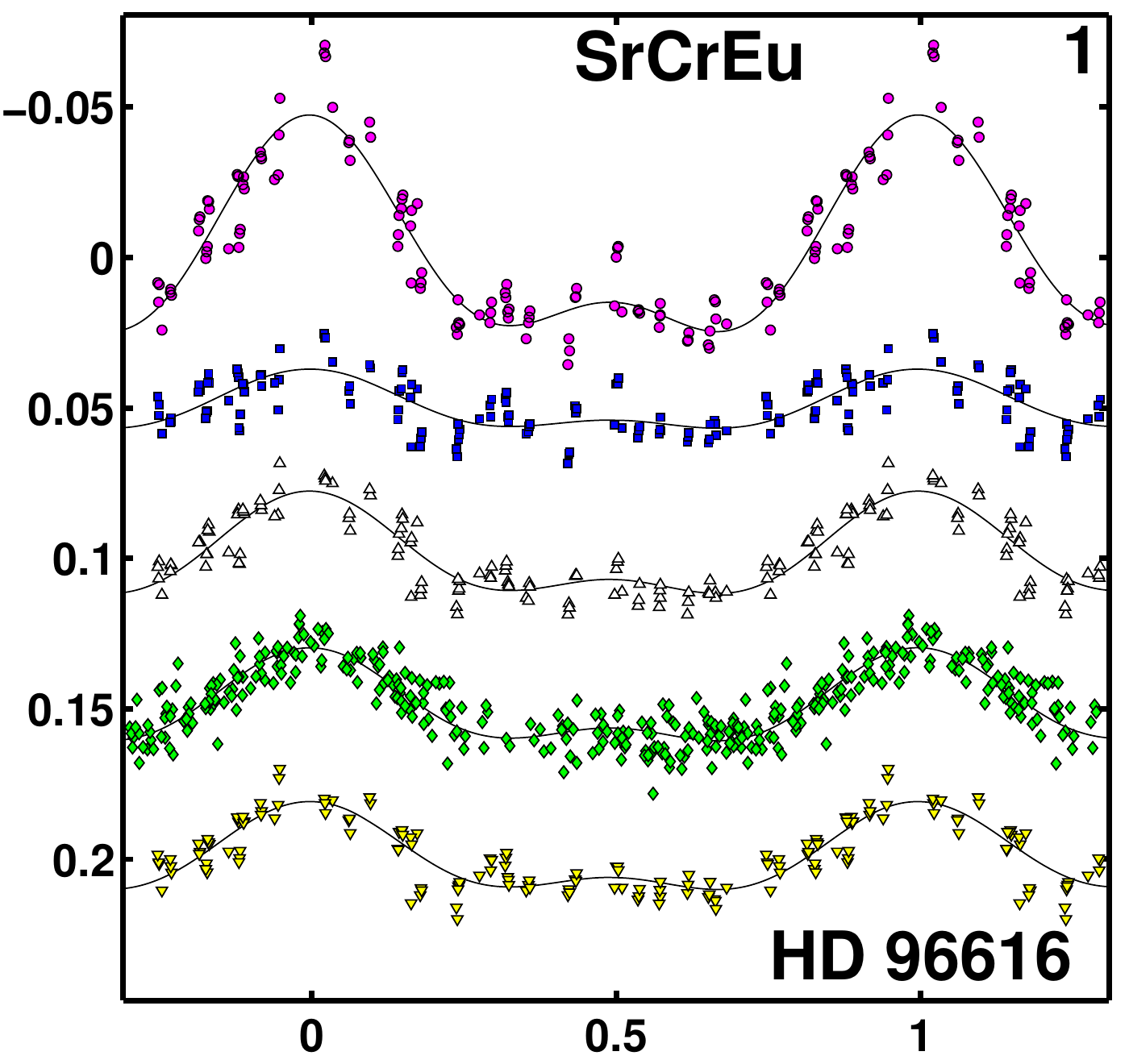}}
\resizebox{0.305\hsize}{0.2\vsize}{\includegraphics{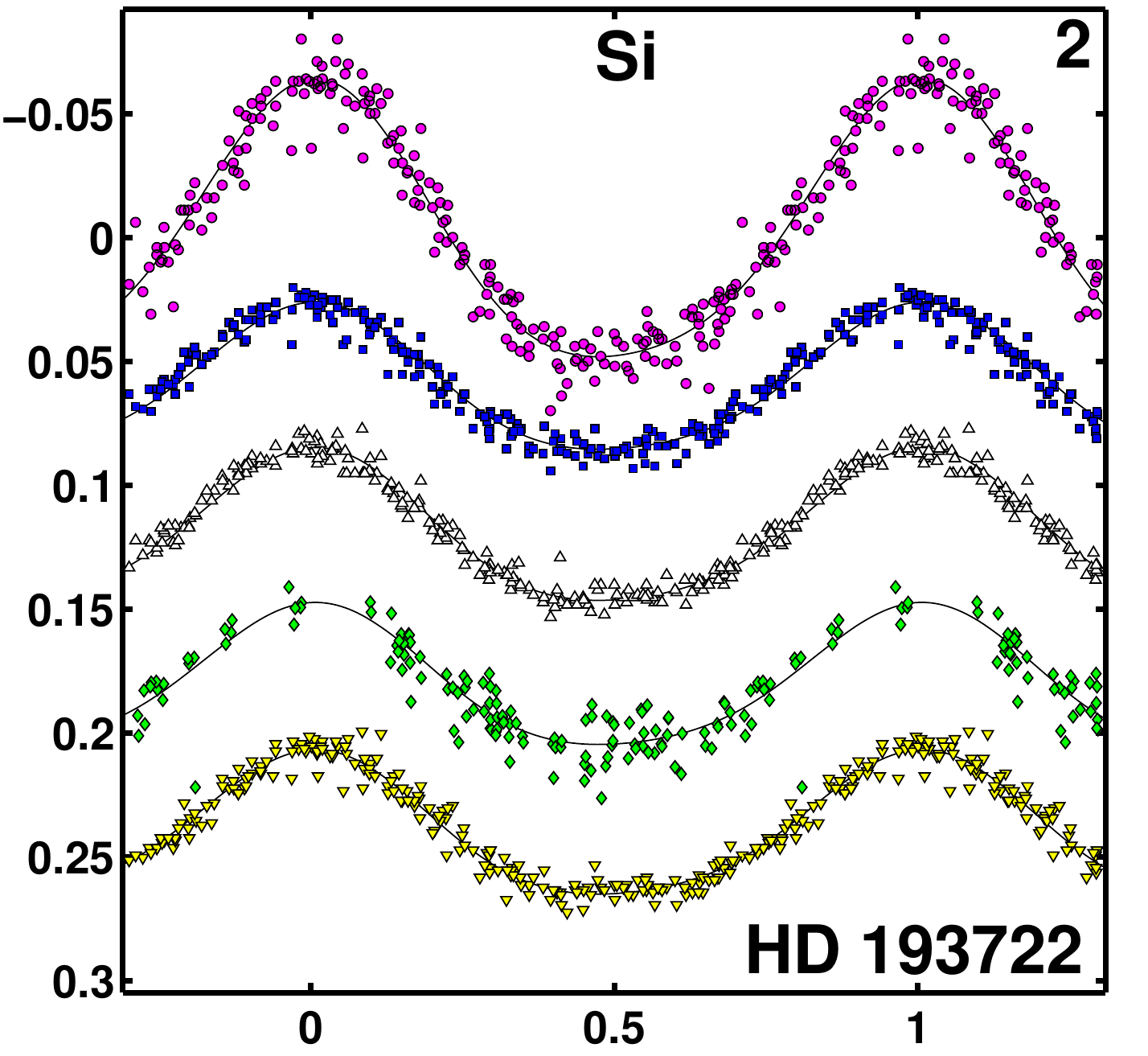}}
\resizebox{0.3\hsize}{0.2\vsize}{\includegraphics{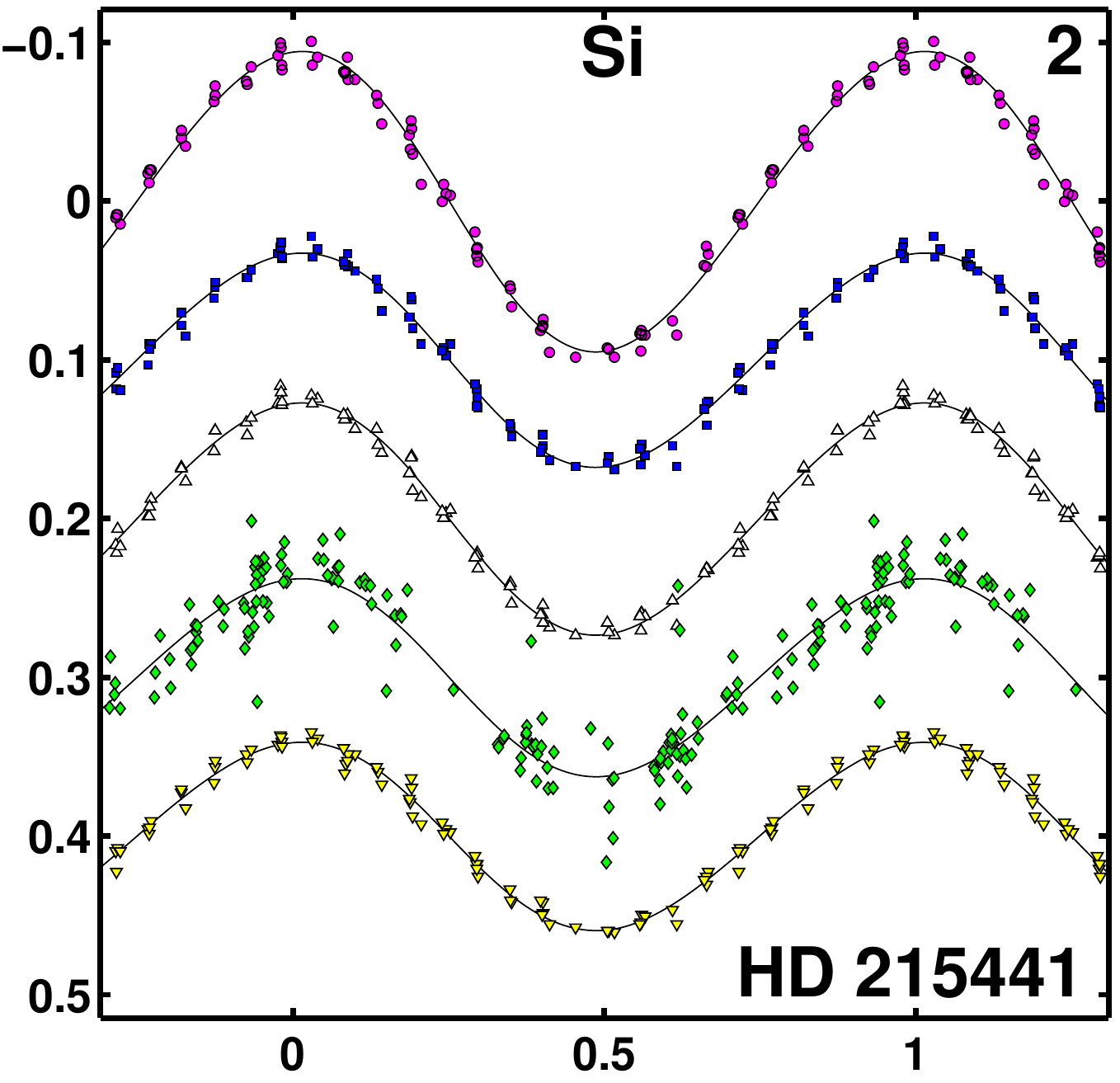}}
\resizebox{0.295\hsize}{0.2\vsize}{\includegraphics{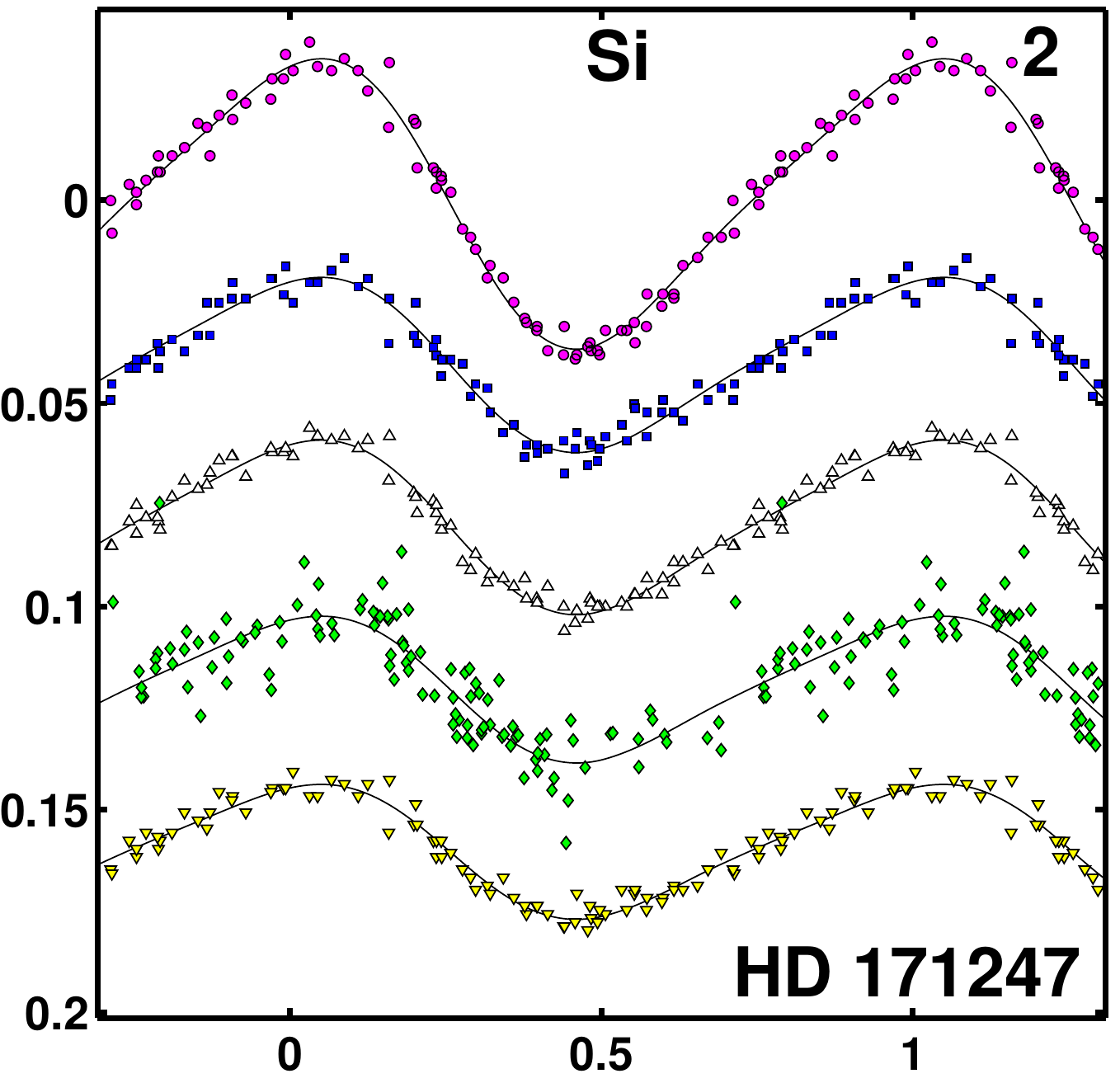}}
\resizebox{0.3\hsize}{0.2\vsize}{\includegraphics{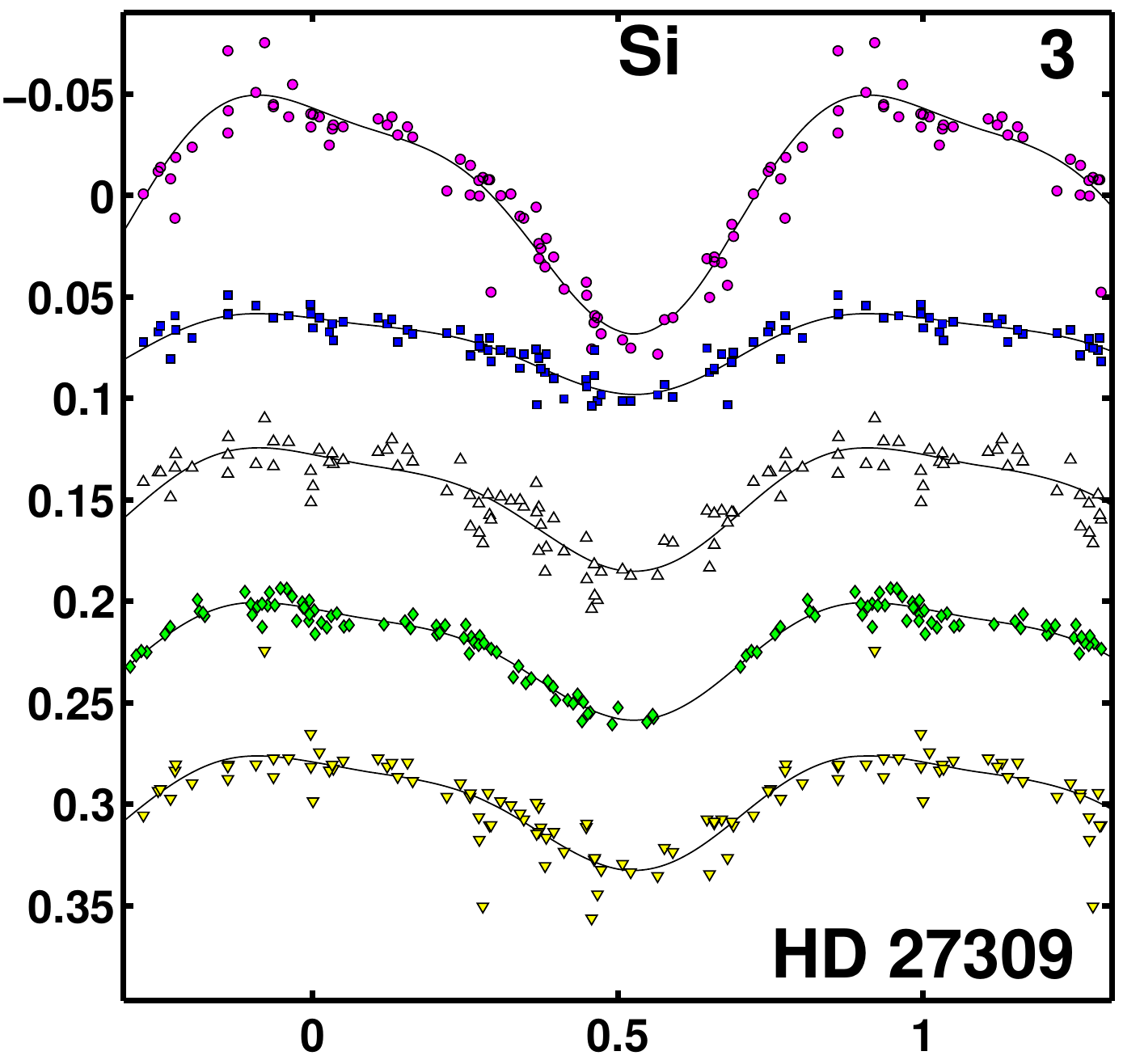}}
\resizebox{0.3\hsize}{0.2\vsize}{\includegraphics{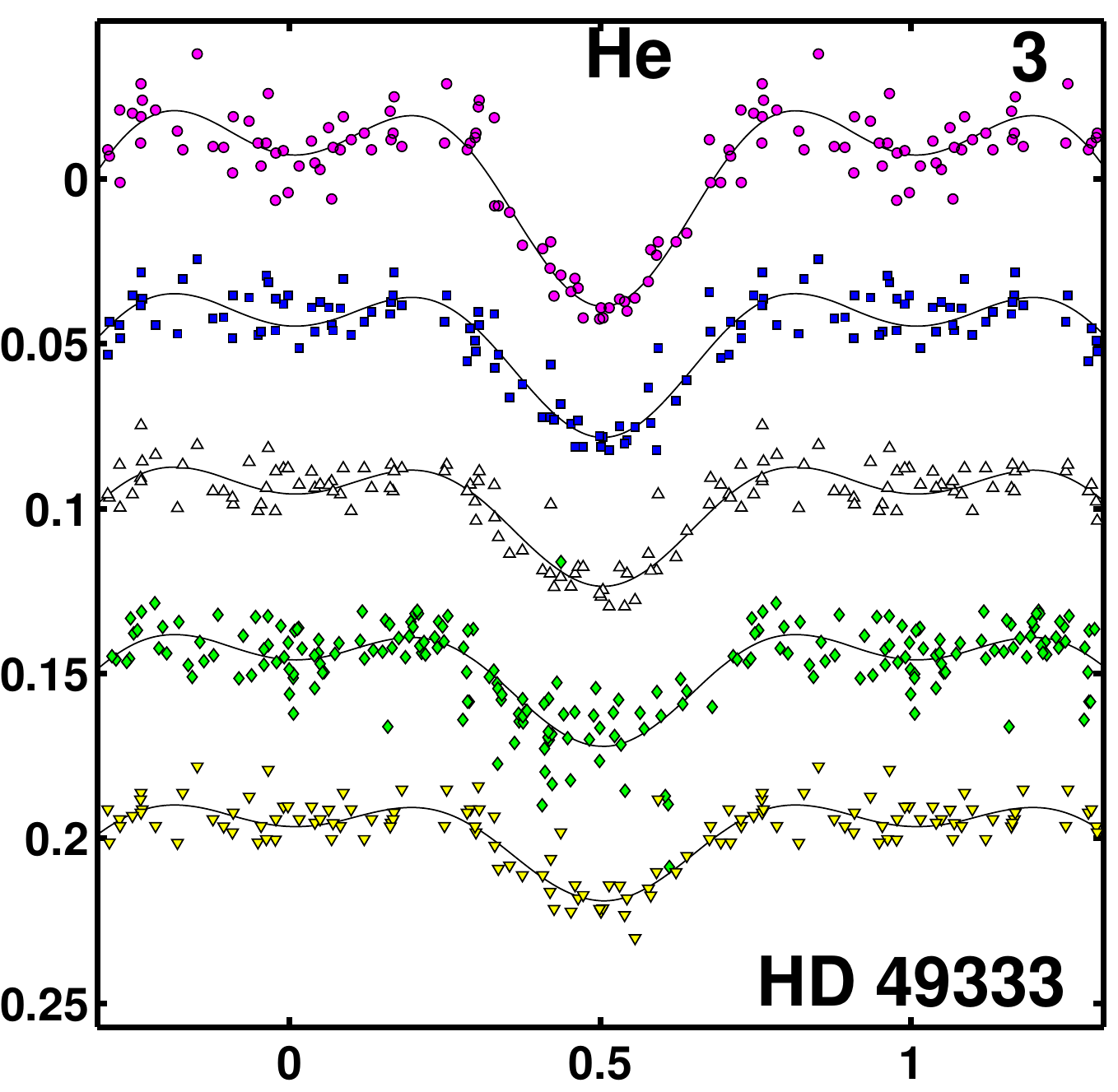}}
\resizebox{0.3\hsize}{0.2\vsize}{\includegraphics{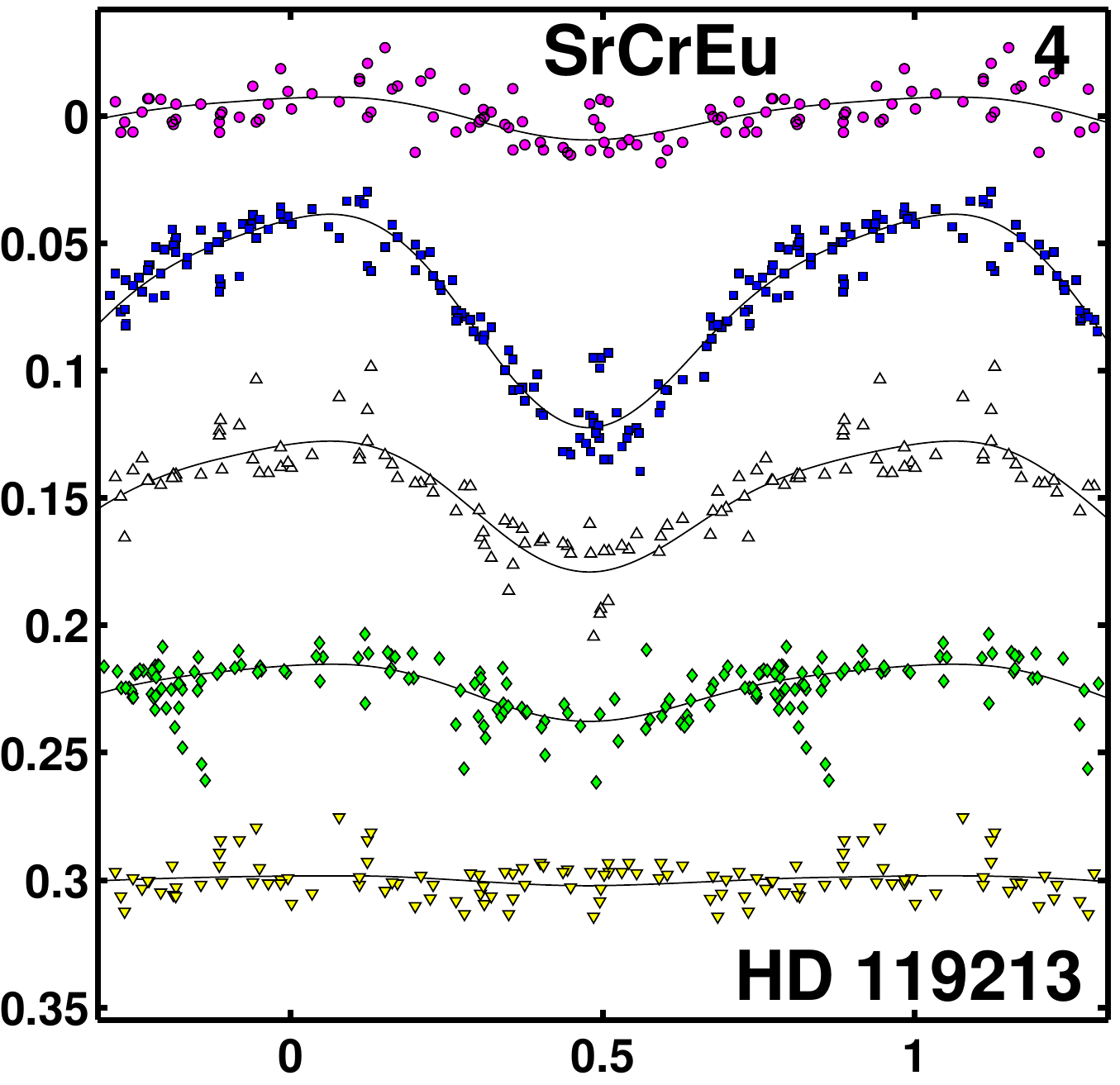}}
\resizebox{0.3\hsize}{0.2\vsize}{\includegraphics{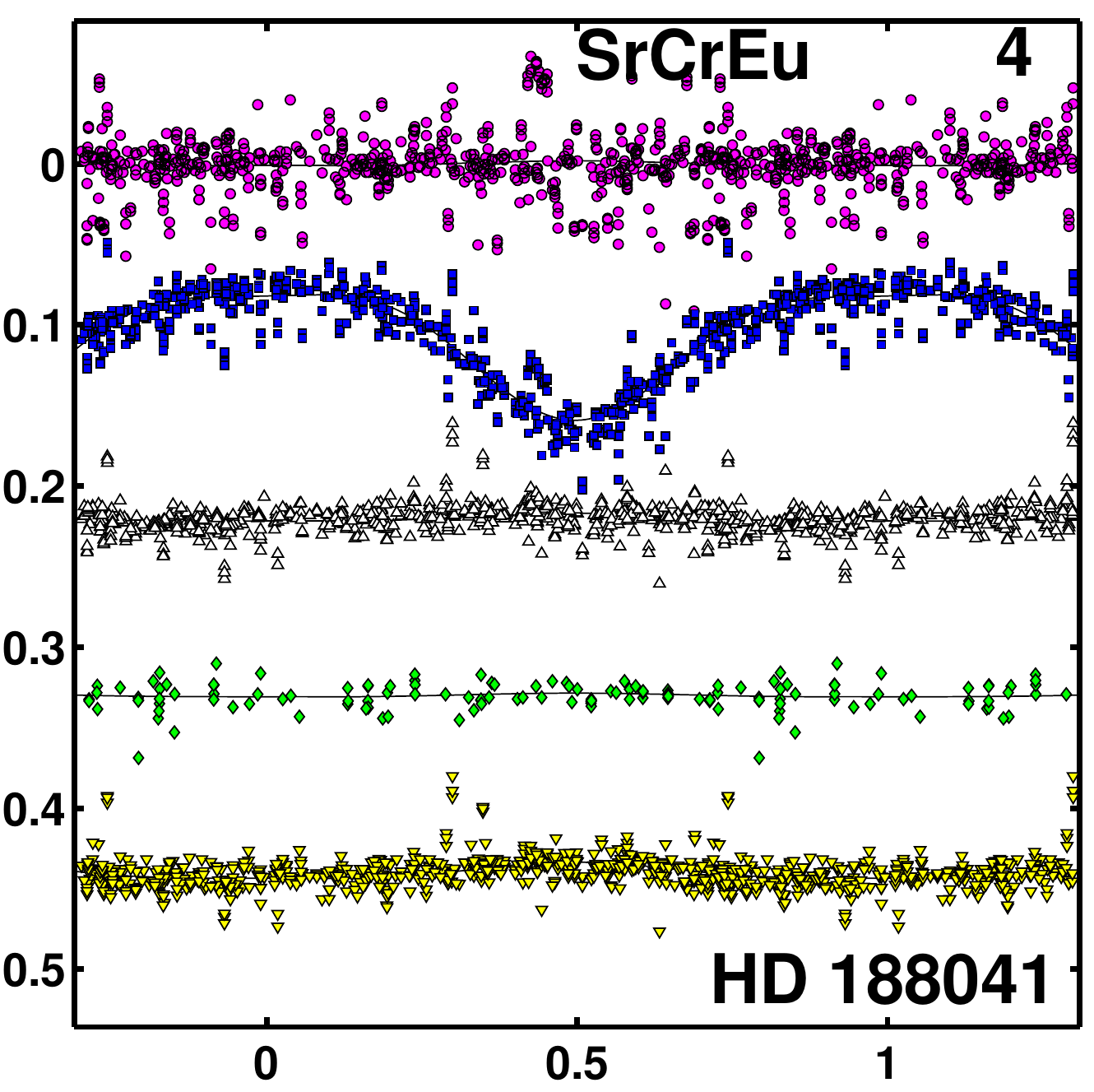}}
\resizebox{0.3\hsize}{0.2\vsize}{\includegraphics{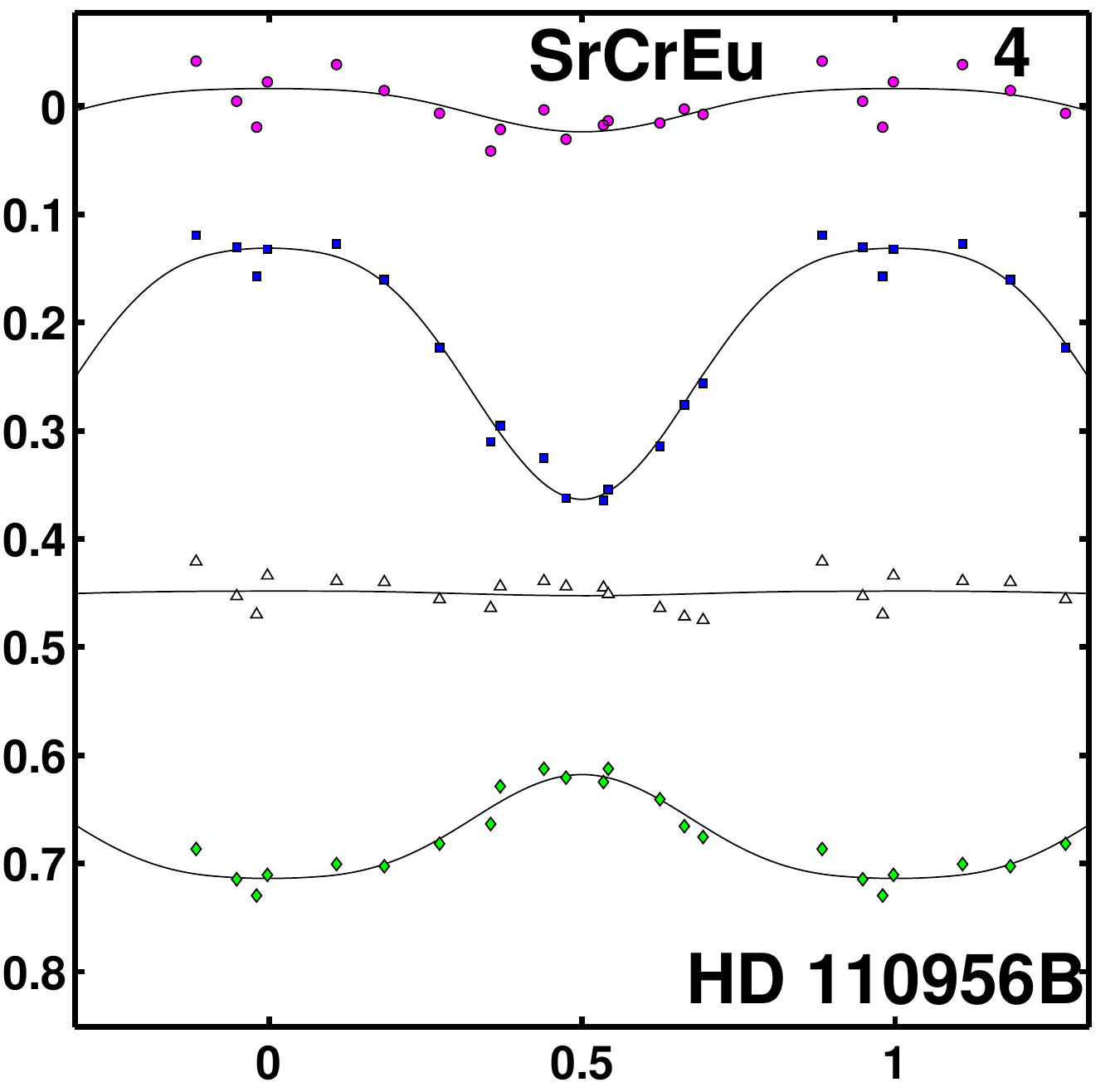}}
\resizebox{0.3\hsize}{0.2\vsize}{\includegraphics{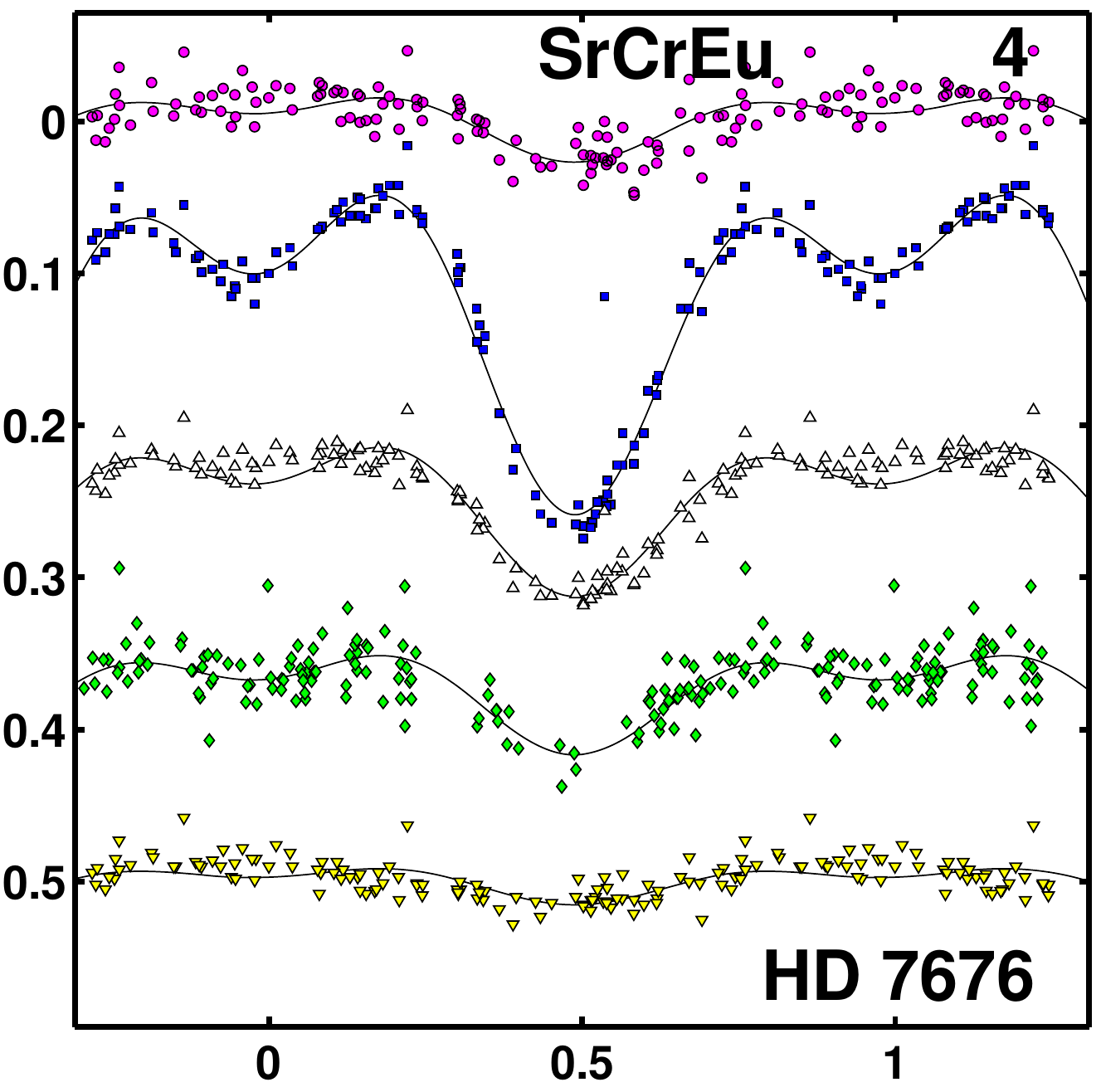}}
\caption[]{Light curves of 12 out of the 19 photometrically simply
behaving mCP stars. In each panel there are displayed $u,v,b,H_p,y$
light curves, the name of the star, its CP spectral type and the
number of the photometric group.} \label{obr}
\end{figure}

Out of the 85 stars with known ephemeris and the $uvby$ and $H_{\rm
p}$ photometries contained in the {\it On-line database of
photometric observations of mCP stars} (Mikul\'a\v{s}ek {\it et
al.}, 2007b) only 19 mCP stars satisfied the above-mentioned
criterion of the similarity: two He-weak, twelve Si, and five
SrCrEu-type stars. It is rather surprising that shapes of their LCs
are very diverse (see Fig.\,\ref{obr}), what can be also documented
by the following quantification. We assume the phase function
$f_1(\varphi)$ in the form:
\begin{eqnarray}
f_1(\varphi)=\sqrt{1-a_{1}^2-a_{2}^2}\
\cos(2\pi\varphi)+a_{1}\cos(4\pi\varphi)
+a_{2}\sin(4\pi\varphi).\label{a12}
\end{eqnarray}
If the timing of the linear ephemeris is set so that the term with
$\sin(2\pi\varphi)$ is zero, then the basic extrema of the LCs occur
near phases 0 and 0.5. The variable part of the LCs of a PSB star is
then characterized by a set of amplitudes $A_{1c}$ which can be
arranged into a vector
$\overrightarrow{A}=[A_{1u},A_{1v},A_{1b},A_{1H_{\rm p}},A_{1y}]$,
and by the two parameters $a_1,\,a_2$ from equation \ref{a12}. A
light curve characterized by $a_{1}>0$ has a sharper maximum at the
phase~0 and a flatter minimum (see e.\,g. HD\,197322) and vice versa
for $a_{1}<0$. The parameter $a_2$ expresses the measure of
asymmetry of an LC. The larger $|a_2|$ the larger asymmetry. The
relation between parameters $a_1,\,a_2$ is depicted on
Fig.\,\ref{a1a2}~left.

\begin{figure}[t]
\centerline{\includegraphics[width=0.8\textwidth]{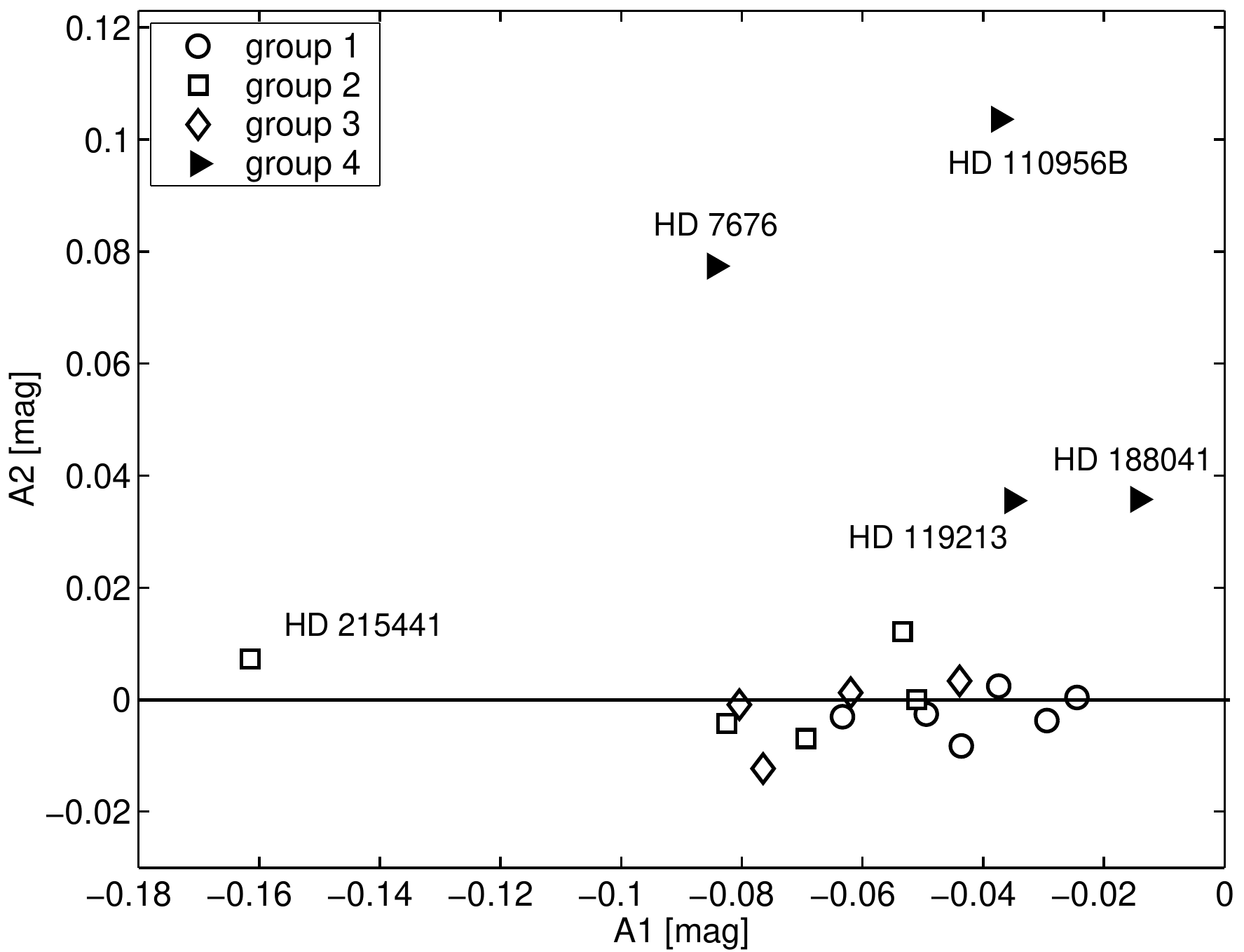}}
\caption{The relation between the effective amplitudes $A_1$ and
$A_2$, corresponding to $\protect\overrightarrow{B_1},
\protect\overrightarrow{B_2}$. The late mCP stars (gr. 4) strongly
differ from other PSBS.} \label{acko12}
\end{figure}

The components of the amplitude vectors $\overrightarrow{A}$ of a
PSB star are not independent. They can be expressed by a linear
combination of two basic normalized vectors $\overrightarrow{B_1}$,
$\overrightarrow{B_2}$ (courses of $B_{1,2}(\lambda)$ -- see
Fig.\,\ref{a1a2}~right, where the wavelength $\lambda$ denotes the
maximum transmissivity of a photometric filter concerned) with
coefficients $A_1,\,A_2$, namely
$\overrightarrow{A}=A_1\,\overrightarrow{B_1}+A_2\,\overrightarrow{B_2}$.
For the relationship of parameters $A_1$ and $A_2$ see
Fig.\,\ref{acko12}. This indicates that there are at least two
different mechanisms responsible for the contrast of a photometric
spot in respect to the surrounding stellar surface. Consequently,
there exist at least two types of photometric spots on the surface
of an mCP star differing in their colour grades. These spots can of
course positionally coincide what can either strengthen or suppress
their resulting contrast in particular photometric colours.

The first mechanism which is obviously effective in all types of mCP
stars creates bright spots in $uvby$  due to redistribution of
energy from the UV region to the optical one originating at least
partly in bound-free transitions of overabundant ions. In hot CP
stars the most important elements seems to be silicon, iron and
helium. The symptomatic feature of this mechanism is the monotonous
decline of the amplitude of light variations $B_2$ with the
increasing effective wavelength as shown on Fig.\,\ref{a1a2}~right.
The effect is well reproduced by our models (Krti\v{c}ka {\it et
al.} 2008, these proceedings).

The second mechanism, manifesting itself through strong variations
of the $c_1$ index, characterizing the height of the Balmer jump, is
present only in the coolest mCP stars. The photometric spots caused
by this mechanism are markedly dark in the $v$ colour, in which the
largest light variations are observed. We speculate that the strong
opacity caused by various overabundant elements in a cool atmosphere
suppresses the Balmer jump. The purest example of the light
variations controlled by this mechanism is HD\,110956B with the
largest amplitude among all mCP stars (see Fig.\,\ref{obr}).

\section{Parametrization and classification of PSB stars}

The set of the $uvby$ and $H_{\rm p}$ light curves of each of the PSB
 stars can be satisfactorily well described by the
quaternion of parameters: $\{a_1,a_2,A_1,A_2\}$.

According to these parameters we sort out all the PSB stars into four
groups.
\begin{itemize}
    \item {\textbf{Group 1}: $A_2\cong0,\ a_1\geq0.24$, double wave
    LCs with two unequally prominent bright spot centered on the phases
    $\varphi=0$~and 0.5. Example: HD\,177410.}
    \item {\textbf{Group 2}: $A_2\cong0,\ |a_1|\leq0.24$, single wave
    LCs. Example: HD\,215441.}
    \item {\textbf{Group 3}: $A_2\cong0,\ a_1\leq-0.24$, double wave
    LCs with two unequally prominent bright spots centered on the
    vicinity of $\varphi=0$. Example: HD\,49333.}
    \item {\textbf{Group 4}: $A_2\neq0$, cool CP stars with the
    largest variations in $v$ caused by a dominating dark spot (see
    HD\,110956B) with a minimum near $\varphi=0.5$. Besides this,
    another dark, as in HD\,7676, or a bright spot, as in HD\,119213
    can be present on the surface.}
\end{itemize}

\section{Conclusions}

All the stars with asymmetric LCs (e.g. HD\,171247), as well as all
the stars with doublewave LCs (groups 1 and 3) must be left out from
the short list of the 19 candidates which LCs can be explained by
the simple one-spot model. Similarly, all the stars of the group 2
with $a_1<0$ must be deleted as they have two or more
undistinguished bright spots centered on the $\varphi=0$. Thus out
of the stars in the groups 1, 2 and 3 only the Si-star HD\,193732 is
the candidate to be explained by the simple one-spot model. Our
analysis of the light curves produced by the one-spot model shows
that for $a_1<0.15$ the spot covers almost the whole hemisphere of a
star.

HD\,188041 and HD\,110956 are the candidates in the group 4.
However, it should be noted, that the dominant photometric spot can,
due to geometry, suppress the visibility of another spot.

We conclude that the presence of two or more photometric spots on
the mCP stars is quite typical. It seems that the photometrically
simply behaving stars do not differ from the others and their
belonging to the privilege group is more likely only the consequence
of the distributions of spots on the surface with respect to the
rotational axis.

\acknowledgements This work was supported by grants GA\,\v{C}R
205/06/0217, VEGA 2/6036/6, and MVTS \v{C}RSR 10/15 and 01506.


\end{document}